\documentclass[12pt,a4paper]{article}
\usepackage[utf8]{inputenc}
\usepackage{amsmath}
\usepackage{amsfonts}
\usepackage{amssymb}
\usepackage{mathtools}
\usepackage{graphicx}
\usepackage{cite}
\usepackage[capitalise]{cleveref}
\newcommand{\bec}{\begin{center}}
\newcommand{\ec}{\end{center}}

\graphicspath{ {./figures/} }
\usepackage[left=2cm,right=2cm,top=2cm,bottom=2cm]{geometry}
\crefformat{section}{#2section~#1#3}
\author{T.V. Obikhod, Ie.A. Petrenko}
\title{\bf{Studying the VBF Hjj and Hjjj Higgs Boson Production in QCD and EFT Theory}}
\date{%
    {\it Institute for Nuclear Research NAS of Ukraine, Kyiv 03028, Ukraine}\\%
    \today
}
\begin{document}

\maketitle

\section{Abstract}

	The discovery of the Higgs boson made it possible not only to study its 
physical properties and update its search strategies, but also to apply 
new theoretical constructs to explain its properties. Within the framework 
of the VBF Higgs boson production process in association with two, Hjj and three jets, Hjjj, 
we modeled its kinematic properties and calculated its production cross-sections. 
The calculations were carried out within two theories, QCD and EFT for both SM processes 
and for the MSSM and NMSSM models within the framework of the BSM physics searches. 

\section{Introduction}
The advent of the LHC has transformed hadron physics into a high-precision
field, where large p$_T$ events at large angles are not uncommon. The search for
physics beyond the Standard Model (SM) implies not only the inclusion of higher-order
corrections to SM processes, but also a change in the search strategy based on
theoretical predictions \cite{Alwall2014}. At the same time, it is necessary to have
universal and accurate simulations for different models, including not only QCD
theory but also others, such as the effective field theory (EFT) \cite{Dong2023}.

	The Higgs boson signal in 2012 was just the beginning of studying its
properties: how strongly it interacts with other particles, to see production and decay
modes of the Higgs boson connected with the information on its couplings to
elementary particles. The study of the vector-boson fusion (VBF) Higgs boson
production processes is related to the accuracy of the electroweak coupling constant
measurement, which is necessary to test the mechanism of spontaneous electroweak
symmetry breaking. Therefore, it is extremely important to separate VBF events from
other background processes, which are achieved by the signature of the VBF process:
the Higgs boson is selected in association with two jets that are strongly separated in
rapidity and form a two-jet system with a high invariant mass to suppress the
contribution of the s-channel. Perturbative next-to-leading order (NLO) corrections to
the QCD Higgs boson production event are considered, in which the two hardest jets
have a rapidity of less than 4.5, which guarantees their detection in the opposite
hemispheres \cite{cruzmartinez2018nnlocorrectionsvbfhiggs}.

	We also study Higgs boson production via VBF in a three-jet association to
match NLO-QCD calculations with the parton shower program. To describe the jet
activity properties in VBF reactions, additional jets are used to suppress QCD
backgrounds by central jet vetoes. Uncertainties due to parton shower effects are
moderate for the third-jet distributions, in contrast to calculations for Higgs
production in a two-jet association 
\cite{nguyen2014mixingprouhetthuemorsesequencesrademacher}.

The paper contains six sections. Section 3 presents calculations of the VBF
Higgs boson production cross sections under the corresponding kinematic and mass
constraints on the jets for three different models, SM, MSSM, and NMSSM. The calculations for the same processes within the EFT theory are given in section 4. The calculations for three-jet events are being reviewed in section 5. In the section 6 the obtained results are analyzed, the order of magnitude of the obtained cross-sections of the processes is compared and the best model 
within the QCD approach is selected. The results of the EFT and QCD
approximations are considered and corresponding conclusions are given.

\section{Studying of VBF Hjj process within the framework of QCD theory}

Understanding the nature of the Higgs boson is associated with its
couplings to fermions and gauge bosons \cite{PhysRevD.70.113009}. 
Such coupling measurements are provided by the VBF processes \cite{EBOLI2000147}, 
where the Higgs boson is produced via quark-scattering mediated 
by weak gauge boson exchange, $pp \to Hjj$, Fig.1

\begin{center}
\includegraphics[width=0.17\textwidth]{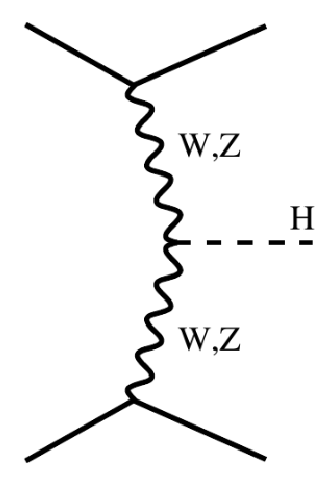}\\
\emph{\textbf{Fig.1}}{\emph{Illustration of the VBF $Hjj$ process.}}
\end{center}
	
	We present a VBF simulation of the Higgs boson production in association 
with two jets obtained at the LHC. The \texttt{\detokenize{MadGraph5_aMC@NLO}} 
computer program \cite{Alwall2014} was used to calculate the NLO-QCD corrections. 
The efficiently suppressed QCD backgrounds with large cross sections to the $Hjj$ 
production process are significant advantages for the accuracy of computer 
modeling of the presented process. 
	
	We carried out modeling of the VBF Higgs boson production cross sections 
with appropriate kinematic restrictions \cite{cruzmartinez2018nnlocorrectionsvbfhiggs}, 
presented in Table 1. 

\vspace{3 mm}
\bec
\emph{\textbf{Table 1.}} {\it  Parameter restrictions for simulations of $Hjj$ production}
\ec
\bec
\begin{tabular}{|c|c|} \hline 
minimum $p_T$ & 20 GeV \\\hline
fixed ren scale & \\
fixed fact scale for pdf1 & 40.21\\
fixed fact scale for pdf2 &  \\\hline
 & 400 GeV \\
min invariant mass of a jet pair & 500 GeV \\
 & 600 GeV   \\\hline
minimum $p_T$ for &  \\
at least one heavy final state & 20 GeV  \\\hline
 minimum $p_T$ for at least one jet  & 20 GeV \\ \hline
 minimum $p_T$ for the leading jet & 25 GeV \\
 minimum $p_T$ for the second jet & 20 GeV \\
 minimum $p_T$ for the third jet & 20 GeV  \\\hline
\end{tabular}
\ec

We carried out process modeling within the framework of QCD with and without \\
\texttt{\detokenize{NNPDF30_nnlo_as_0118}} pdf function. As a part of 
the search for physics beyond the SM, we compared the modeling of 
SM, MSSM, and NMSSM processes in the framework of $Hjj$ cross-section 
calculations, presented in Table 2 and Table 3

\vspace{3 mm}
\bec
\emph{\textbf{Table 2.}} {\it  Production cross sections for $p p \to h j j$ 
process within three models with pdf function}
\ec
\bec
\begin{tabular}{|c|c|c|c|c|} 
\hline 
$M_{mjj}$ & Model & 13TeV & 14TeV & 100TeV \\
$W^\pm Z$ & & & & \\ \hline

400 
 & SM    & \shortstack{$1.735$ \\ $\pm 0.009$}  & \shortstack{$2.001$ \\ $\pm 0.01$}  
& \shortstack{$28.13$ \\ $\pm 0.092$}  \\ \cline{2-5}
 & mssm  & \shortstack{$1.664$ \\ $\pm 0.0095$} & \shortstack{$1.911$ \\ $\pm 0.0085$} 
& \shortstack{$26.97$ \\ $\pm 0.086$} \\ \cline{2-5}
 & nmssm & \shortstack{$2.16$ \\ $\pm 0.012$}   & \shortstack{$2.492$ \\ $\pm 0.012$}  
& \shortstack{$35.16$ \\ $\pm 0.12$}  \\ \hline

500 
 & SM    & \shortstack{$1.501$ \\ $\pm 0.006$}  & \shortstack{$1.709$ \\ $\pm 0.01$}  
& \shortstack{--- \\ ---} \\ \cline{2-5}
 & mssm  & \shortstack{$1.431$ \\ $\pm 0.007$}  & \shortstack{$1.64$ \\ $\pm 0.011$}  
& \shortstack{--- \\ ---} \\ \cline{2-5}
 & nmssm & \shortstack{$1.863$ \\ $\pm 0.008$}  & \shortstack{$2.132$ \\ $\pm 0.014$}  
& \shortstack{--- \\ ---} \\ \hline

600 
 & SM    & \shortstack{$1.285$ \\ $\pm 0.007$}  & \shortstack{$1.491$ \\ $\pm 0.012$}  
& \shortstack{--- \\ ---} \\ \cline{2-5}
 & mssm  & \shortstack{$1.228$ \\ $\pm 0.007$}  & \shortstack{$1.426$ \\ $\pm 0.01$}   
& \shortstack{--- \\ ---} \\ \cline{2-5}
 & nmssm & \shortstack{$1.6$ \\ $\pm 0.009$}    & \shortstack{$1.86$ \\ $\pm 0.012$}   
& \shortstack{--- \\ ---} \\ \hline

\end{tabular}
\ec

\vspace{3 mm}
\bec
\emph{\textbf{Table 3.}} {\it  Production cross sections for $p p \to h j j$ 
process within three models without pdf function}
\ec
\bec
\begin{tabular}{|c|c|c|c|c|} 
\hline 
$M_{mjj}$ & Model & 13TeV & 14TeV & 100TeV \\ \hline

400 
 & SM    & \shortstack{$1.563$ \\ $\pm 0.007$}  & \shortstack{$1.778$ \\ $\pm 0.006$}  
& \shortstack{$23.28$ \\ $\pm 0.077$}  \\ \cline{2-5}
 & mssm  & \shortstack{$1.498$ \\ $\pm 0.007$}  & \shortstack{$1.708$ \\ $\pm 0.006$}  
& \shortstack{$22.16$ \\ $\pm 0.074$}  \\ \cline{2-5}
 & nmssm & \shortstack{$1.948$ \\ $\pm 0.009$}  & \shortstack{$2.228$ \\ $\pm 0.011$}  
& \shortstack{$28.96$ \\ $\pm 0.1$}    \\ \hline

500 
 & SM    & \shortstack{$1.362$ \\ $\pm 0.0054$}  & \shortstack{$1.542$ \\ $\pm 0.009$}  
& \shortstack{--- \\ ---} \\ \cline{2-5}
 & mssm  & \shortstack{$1.302$ \\ $\pm 0.007$}   & \shortstack{$1.48$ \\ $\pm 0.01$}   
& \shortstack{--- \\ ---} \\ \cline{2-5}
 & nmssm & \shortstack{$1.692$ \\ $\pm 0.008$}   & \shortstack{$1.915$ \\ $\pm 0.01$}  
& \shortstack{--- \\ ---} \\ \hline

600 
 & SM    & \shortstack{$1.168$ \\ $\pm 0.006$}  & \shortstack{$1.348$ \\ $\pm 0.008$}  
& \shortstack{--- \\ ---} \\ \cline{2-5}
 & mssm  & \shortstack{$1.119$ \\ $\pm 0.006$}  & \shortstack{$1.293$ \\ $\pm 0.009$}  
& \shortstack{--- \\ ---} \\ \cline{2-5}
 & nmssm & \shortstack{$1.456$ \\ $\pm 0.007$}  & \shortstack{$1.683$ \\ $\pm 0.01$}   
& \shortstack{--- \\ ---} \\ \hline

\end{tabular}

\ec

	Comparison of the obtained results presented in Table 2 and Table 3  
shows the highest value of the production cross-section of $Hjj$ processes 
with pdf function and for \texttt{nmssm} model with a limitation on the mass 
of the two-jet process in the region of 400 GeV. Moreover, the priority 
of this model is seen for all energies, 13, 14 and 100 TeV. As for 
calculations within this model for other invariant masses (500 GeV, 600 GeV), 
the software we use does not allow us to perform these calculations 
for the energy at the 100 TeV collider.

To understand the kinematics of the process and to ensure the correctness 
of the chosen kinematic constraints, simulations of the distribution 
of transverse momentum and rapidity were carried out in Fig. 2, 3.

\begin{center}
\includegraphics[width=0.5\textwidth]{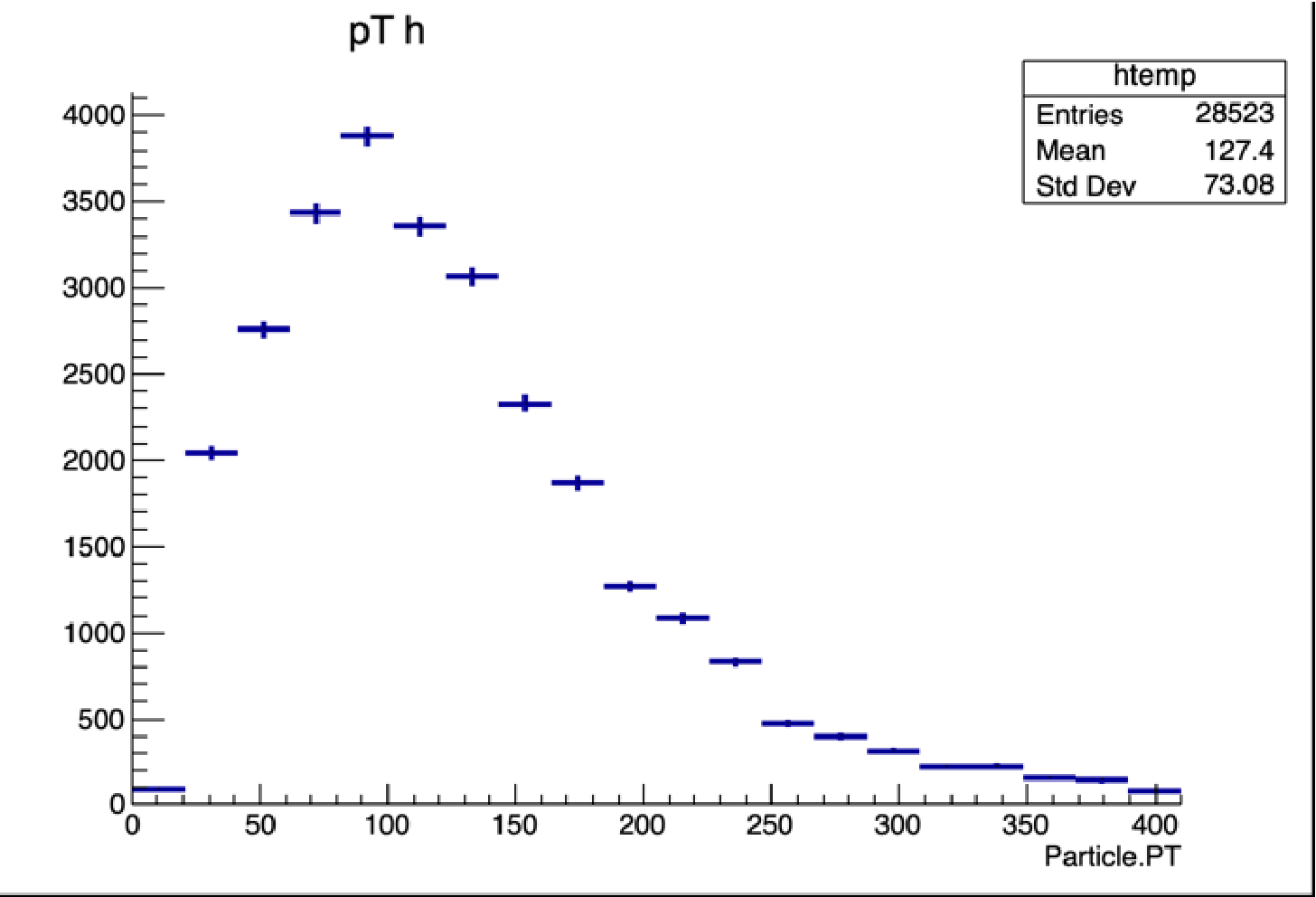}\\
\emph{\textbf{Fig.2}}{\emph{Transverse momentum distributions for $p p \to h j j$ process at 14 TeV.}}
\end{center}

\begin{center}
\includegraphics[width=0.5\textwidth]{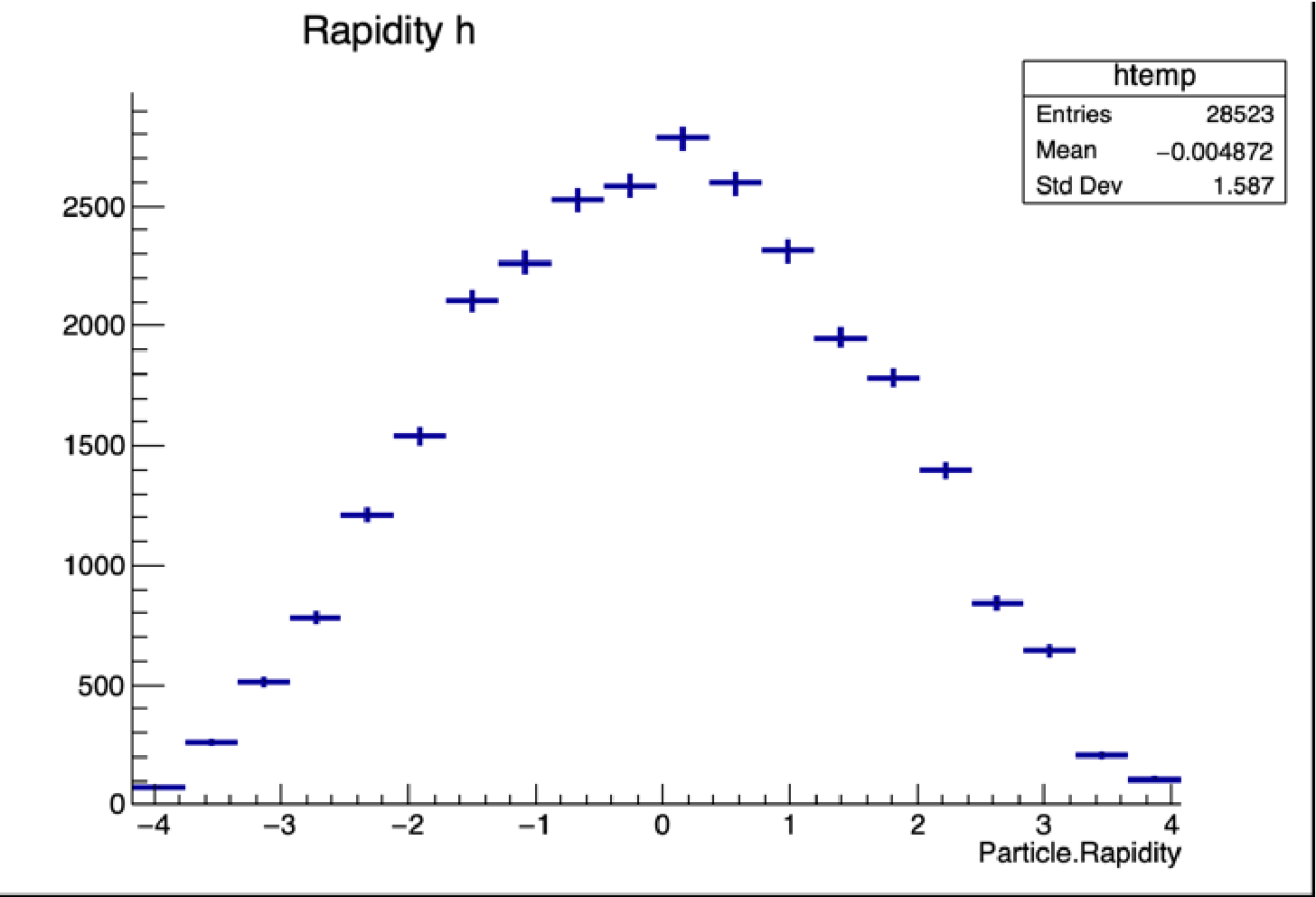}\\
\emph{\textbf{Fig.3}}{\emph{The distribution for the rapidity for $p p \to h j j$ process at 14 TeV.}}
\end{center}

The results of the modeling demonstrate the peripheral nature of the process under 
consideration and the clear dependence of the number of events on the value of the transverse momentum.

\section{Studying of VBF $Hjj$ process in EFT theory}

EFT is an approach in theoretical physics based on the idea of describing physical phenomena for any pre-selected energy scale, from the minimum to the maximum energy scale. This theory is convenient 
for describing various sections of physics, including elementary particle physics. As a more the fundamental theory is lacking it is necessary to resort to an EFT description of a physical system, which has next advantages:

\vspace{-\topsep}	 
\begin{itemize}
\setlength{\parskip}{0pt}
  \setlength{\itemsep}{0pt plus 0.5pt}
  \item EFT radically simplifies calculations by ignoring aspects that are not important 
for the problem at hand;
  \item By extracting the appropriate degrees of freedom, new symmetries may emerge 
that would otherwise remain hidden;
  \item When working with problems that have several different scales, EFTs allow us 
to focus on one scale to perform RG summations of large logarithms;
  \item EFT is convenient because if perturbation theory fails, you can pick up other 
degrees of freedom within the new EFT.
\end{itemize}
\vspace{-\topsep}

Notably, EFT with massive fields finds a broad range of applications in particle physics, including quantum chromodynamics, high spin particles, and dark matter candidates. This method is agnostic to how electroweak symmetry is broken, so the consideration of mass eigenstates allows the formulation of EFT to be much simpler and more convenient for phenomenological applications. In the framework of the Higgs effective field theory \cite{Dong2023} we considered its link to a few scenarios of physics beyond the SM and received the production cross-sections presented in Table 4
	
\vspace{3 mm}
\bec
\emph{\textbf{Table 4.}} {\it  Production cross sections for $p p \to h j j$ process 
with pdf function within EFT theory.}
\ec
\bec
\begin{tabular}{|c|c|c|c|c|} 
\hline 
$M_{mjj}$ & Model & 13TeV & 14TeV & 100TeV \\ \hline

400 
 & SM    & \shortstack{$1.73$ \\ $\pm 0.008$}  & \shortstack{$1.996$ \\ 
$\pm 0.01$}  & \shortstack{$28.25$ \\ $\pm 0.084$}  \\ \cline{2-5}
 & mssm  & \shortstack{$1.663$ \\ $\pm 0.007$}  & \shortstack{$1.917$ \\ 
$\pm 0.009$}  & \shortstack{$26.96$ \\ $\pm 0.083$}  \\ \cline{2-5}
 & nmssm & \shortstack{$2.162$ \\ $\pm 0.009$}  & \shortstack{$2.489$ \\ 
$\pm 0.012$}  & \shortstack{$35.14$ \\ $\pm 0.12$}   \\ \hline

500 
 & SM    & \shortstack{$1.498$ \\ $\pm 0.0065$}  & \shortstack{$1.996$ \\ 
$\pm 0.01$}  & \shortstack{--- \\ ---} \\ \cline{2-5}
 & mssm  & \shortstack{$1.427$ \\ $\pm 0.0063$}  & \shortstack{$1.64$ \\ 
$\pm 0.011$}  & \shortstack{--- \\ ---} \\ \cline{2-5}
 & nmssm & \shortstack{$1.859$ \\ $\pm 0.009$}   & \shortstack{$2.132$ \\ 
$\pm 0.01$}  & \shortstack{--- \\ ---} \\ \hline

600 
 & SM    & \shortstack{$1.498$ \\ $\pm 0.0065$}  & \shortstack{$1.492$ \\ 
$\pm 0.012$}  & \shortstack{--- \\ ---} \\ \cline{2-5}
 & mssm  & \shortstack{$1.225$ \\ $\pm 0.007$}  & \shortstack{$1.424$ \\ 
$\pm 0.01$}   & \shortstack{--- \\ ---} \\ \cline{2-5}
 & nmssm & \shortstack{$1.593$ \\ $\pm 0.009$}  & \shortstack{$1.853$ \\ 
$\pm 0.01$}   & \shortstack{--- \\ ---} \\ \hline

\end{tabular}

\ec

			The analysis of the obtained results shows that the cross-sections 
of Tables 2 and 4 almost coincide for all models and at all levels. However, 
for the SM, a significant increase in the cross-section for the EFT is observed 
at an invariant mass of the two-jet state of 600 GeV at an energy at the collider of 13 TeV.

\section{Studying of VBF $Hjjj$ process within the QCD theory}

The observation of two forward tagging jets in Higgs production via 
the VBF at the LHC is crucial for background suppression. 
In addition to forward tagging jets, the veto of any additional 
jet activity in the central region leads to further suppression 
of QCD backgrounds. If any additional jet with momentum higher 
than the minimum value p$_T$ is detected in addition to the two 
forward tagging jets, the events are discarded \cite{TerranceFigy2008}.

We present a realization of Higgs boson production via VBF in association 
with three jets. We accurately described the properties of the additional 
jet in VBF reactions to suppress QCD backgrounds in Table 1,
\cite{Jager2014}. Due to the low virtuality of the exchanged weak bosons, 
the tagging jets arising from the scattered quarks are located in the front 
and back regions of the detector, while the central region exhibits little 
jet activity due to the t-channel exchange. These features can be used 
to suppress QCD backgrounds with a large cross-section at the LHC. 
Feynman diagrams of such processes are shown in Fig. 4.

\begin{center}
\includegraphics[width=0.3\textwidth]{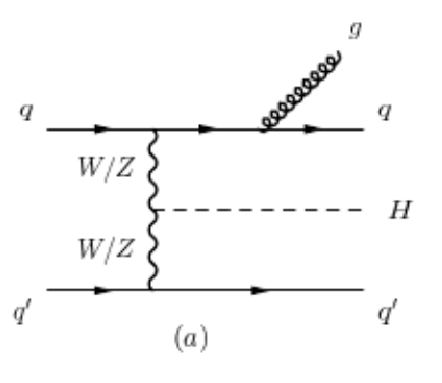}\\
\includegraphics[width=0.3\textwidth]{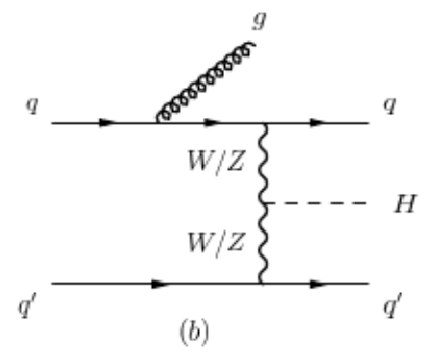}\\
\emph{\textbf{Fig.4}}{\emph{Tree-level diagrams for VBF $Hjjj$ production \cite{Jager2014}.}}
\end{center}

The inclusion of the constraints presented in Table 1 and the use of the 
\texttt{\detokenize{MadGraph5_aMC@NLO}} computer program makes it possible 
to calculate the cross sections of the three-jet processes, Table 5.

\begin{table*}
    \centering
    \emph{\textbf{Table 5.}} {\it  Production cross sections 
for $p p \to h jjj$ process with pdf function within three models with pdf function.} \\
    \vspace{3 mm}
\begin{tabular}{|c|c|c|c|c|} 
\hline 
$M_{mjj}$ & Model & 13TeV & 14TeV & 100TeV \\ \hline

400 
 & SM    & \shortstack{$0.0056$ \\ $\pm 3.6 \times 10^{-5}$} 
 & \shortstack{$0.0071$ \\ $\pm 4.8 \times 10^{-5}$}  
& \shortstack{$0.42$ \\ $\pm 0.0018$}  \\ \cline{2-5}
 & mssm  & \shortstack{$0.0053$ \\ $\pm 3.6 \times 10^{-5}$}  
& \shortstack{$0.0066$ \\ $\pm 4.7 \times 10^{-5}$}  
& \shortstack{$0.398$ \\ $\pm 0.002$}  \\ \cline{2-5}
 & nmssm & \shortstack{$0.007$ \\ $\pm 4.8 \times 10^{-5}$}  
& \shortstack{$0.0087$ \\ $\pm 6.4 \times 10^{-5}$}  & \shortstack{$0.53$ 
\\ $\pm 0.002$}   \\ \hline

500
 & SM    & \shortstack{$0.0056$ \\ $\pm 3.6 \times 10^{-5}$}  & \shortstack{$0.0027$ \\ 
$\pm 2.0 \times 10^{-5}$}  & \shortstack{$0.22$ \\ $\pm 9.0 \times 10^{-4}$} \\ \cline{2-5}
 & mssm  & \shortstack{$0.002$ \\ $\pm 1.3 \times 10^{-5}$}  & \shortstack{$0.0026$ \\
 $\pm 1.8 \times 10^{-5}$}  & \shortstack{$0.21$ \\ $\pm 1.0 \times 10^{-4}$} \\ \cline{2-5}
 & nmssm & \shortstack{$0.0027$ \\ $\pm 1.6 \times 10^{-5}$}  & \shortstack{$0.0034$ \\
 $\pm 2.6 \times 10^{-5}$}  & \shortstack{$0.28$ \\ $\pm 1.0 \times 10^{-3}$} \\ \hline

600 
 & SM    & \shortstack{$0.0009$ \\ $\pm 6.1 \times 10^{-6}$}  & \shortstack{$0.0012$ \\ 
$\pm 7.2 \times 10^{-6}$}  & \shortstack{$0.12$ \\ $\pm 0.0006$}  \\ \cline{2-5}
 & mssm  & \shortstack{$0.0008$ \\ $\pm 5.8 \times 10^{-6}$}  & \shortstack{$0.0011$ 
\\ $\pm 6.6 \times 10^{-6}$}  & \shortstack{$0.12$ \\ $\pm 0.0005$}  \\ \cline{2-5}
 & nmssm & \shortstack{$0.0011$ \\ $\pm 7.9 \times 10^{-6}$}  & \shortstack{$0.0014$ \\ 
$\pm 8.6 \times 10^{-6}$}  & \shortstack{$0.155$ \\ $\pm 0.0007$}  \\ \hline

\end{tabular}

 \end{table*}	

Comparison of the obtained data for three-jet events with the data in Table 2 
for two-jet processes leads us to the conclusion about the same nature 
of the behavior of the obtained data (with minor exceptions), i.e.:
a decrease in the cross-section of the process with an increase in the invariant mass of the two jets; an increase in the cross-section with an increase in the energy at the collider from 13 to 100 TeV; the largest value of the cross-section for the \texttt{nmssm} model.
We compared our calculations with the data from the articles \cite{Alwall2014,TerranceFigy2008} 
and found a good match in the order of magnitude for both two-jet and three-jet events. Moreover, the cross-section value of three-jet events is two orders of magnitude smaller than that of two-jet events, which confirms the level of influence of three-jet constraints on the cross-section of the process $Hjjj$.

\section{Conclusions}

In the context of studying the properties of the Higgs boson, we presented 
two processes of its VBF formation: two-jet and three-jet. Comparison of 
the obtained results showed that our calculations coincided by order of 
magnitude with the calculations of other scientists for both types of processes. 
In addition, we carried out modeling of the BSM theories and found that 
the largest cross section is for the \texttt{nmssm} model. Comparison with 
the EFT, which is popular recently due to its flexibility 
and universality, led us to the conclusion that the cross sections of the 
Higgs boson production calculated within the framework of this theory are 
larger for the EFT SM model compared to the SM QCD calculations. The 
kinematic distributions constructed by us confirmed a strict dependence 
on the transverse momentum value and indicated a large number of events at large angles.

\end{document}